\font\twelverm = cmr10 scaled\magstep1 \font\tenrm = cmr10
       \font\sevenrm = cmr7
\font\twelvei = cmmi10 scaled\magstep1
       \font\teni = cmmi10 \font\seveni = cmmi7
\font\twelveit = cmti10 scaled\magstep1 
       
\font\twelvesy = cmsy10 scaled\magstep1
       \font\tensy = cmsy10 \font\sevensy = cmsy7
\font\twelvebf = cmbx10 scaled\magstep1 \font\tenbf = cmbx10
       \font\sevenbf = cmbx7
\font\twelvesl = cmsl10 scaled\magstep1
\font\twelvett = cmtt10 scaled\magstep1
%
\textfont0 = \twelverm \scriptfont0 = \twelverm
       \scriptscriptfont0 = \tenrm
       \def\rm{\fam0 \twelverm}
\textfont1 = \twelvei \scriptfont1 = \twelvei
       \scriptscriptfont1 = \teni
       
\textfont2 = \twelvesy \scriptfont2 = \twelvesy
       \scriptscriptfont2 = \tensy
       
\newfam\itfam \def\it{\fam\itfam \twelveit} \textfont\itfam=\twelveit
\newfam\slfam \def\sl{\fam\slfam \twelvesl} \textfont\slfam=\twelvesl
\newfam\bffam \def\bf{\fam\bffam \twelvebf} \textfont\bffam=\twelvebf
       \scriptfont\bffam=\twelvebf \scriptscriptfont\bffam=\tenbf
\newfam\ttfam  \textfont\ttfam=\twelvett
\rm
\hsize=6.5in
\hoffset=.1in
\vsize=9in
\baselineskip=24pt
%
\raggedright  \pretolerance = 800  \tolerance = 1100
\raggedbottom
%
\dimen1=\baselineskip \multiply\dimen1 by 3 \divide\dimen1 by 4
\dimen2=\dimen1 \divide\dimen2 by 2
%
\def\apjsingle{\baselineskip = 14pt
               \parskip=14pt plus 1pt
               \dimen1=\baselineskip \multiply\dimen1 by 3 \divide\dimen1 by 4
               \dimen2=\dimen1 \divide\dimen2 by 2
               \scriptfont0 = \tenrm  \scriptscriptfont0 = \sevenrm
               \scriptfont1 = \teni  \scriptscriptfont1 = \seveni
               \scriptfont2 = \tensy \scriptscriptfont2 = \sevensy
               \scriptfont\bffam = \tenbf \scriptscriptfont\bffam = \sevenbf
               \rightskip=0pt  \spaceskip=0pt  \xspaceskip=0pt
               \pretolerance=200  \tolerance=400
              }
%
\nopagenumbers
\headline={\ifnum\pageno=1 \hss\thinspace\hss
     \else\hss\folio\hss \fi}
%

%
\count10 = 0
\def\section#1{\vbox to \dimen1 {\vfill}
    \global\advance\count10 by 1
    \centerline{\expandafter{\number\count10}.\ \bf {#1}}
    \global\count11=96  
    \vskip \dimen1}
%
\def\subsection#1{\global\advance\count11 by 1
    \vskip \parskip  \vskip \dimen2
    \centerline{{\it {\char\number\count11}\/})\ {\it #1}}
    \vskip \dimen2}
%
%
\def\refindent{\advance\leftskip by 24pt \parindent=-24pt}
%

%

%

%

%

%

%

%

%

%

%

\def\lsim{\raise0.3ex\hbox{$<$}\kern-0.75em{\lower0.65ex\hbox{$\sim$}}}
\def\gsim{\raise0.3ex\hbox{$>$}\kern-0.75em{\lower0.65ex\hbox{$\sim$}}}
\baselineskip= 20pt
\def \yskip{\penalty-50\vskip3pt plus 3pt minus 2pt}
\def \pp{\par \yskip \noindent \hangindent .4in \hangafter 1}
\def \abc#1#2#3#4 {\pp#1, {\sl#2}, {\bf#3}, #4}
\def \blank {\lower 5pt\hbox to 0.75in{\hrulefill}}
\def \lam {$\lambda$}

\def \lam{$\lambda$}

%
%
\apjsingle
\def \yskip{\penalty-50\vskip3pt plus 3pt minus 2pt}
\def \pp{\par \yskip \noindent \hangindent .4in \hangafter 1}
\def \abc#1#2#3#4 {\pp#1, {\sl#2}, {\bf#3}, #4}
\def \blank {\lower 5pt\hbox to 0.75in{\hrulefill}}

\def \mdot {\rm{M}$_\odot$~yr$^{-1}$}
\def \lam {$\lambda$}

\def \lam{$\lambda$}

\def \apj{{\it Ap.J.,}}
\def \aj{{\it A.J.,}}
\def \aa{{\it A.\& A.,}}
\def \nat{{\it Nature,}}
\def \mn{{\it MNRAS,}}
\tolerance=10000
\centerline{\bf From the Owl to the Eskimo:}
\centerline{\bf The Radiation-Gasdynamics of Planetary Nebulae IV}
\vskip  1cm
\centerline{Adam Frank$^1$ and Garrelt Mellema$^{2,3}$}
\bigskip
\bigskip
\bigskip
\bigskip
\bigskip
\line{$^1$Department of Astronomy, University of Minnesota,
Minneapolis, MN 55455 \hfill}
\line{$^2$Sterrewacht Leiden, P.O. Box 9513, Leiden University,
2300 RA Leiden The Netherlands \hfill}
\line{$^3$Dept.\ of Mathematics, UMIST, P.O. Box 88, Manchester M60
1QD, UK\hfill}
\bigskip
\bigskip
\bigskip
\bigskip
\bigskip
\bigskip
\centerline{\bf Accepted by the Astrophysical Journal}
\bigskip
\bigskip
\bigskip
\bigskip
\bigskip

\vfill\eject

\centerline{\bf Abstract}
We present the results of two-dimensional radiation-gasdynamic
simulations of aspherical Planetary Nebulae (PNe) evolution.  These
simulations were constructed using the Generalized Interacting Stellar
Winds (GISW) scenario of Balick (1987) where a fast, tenuous wind from
the central star expands into a toroidal, slow, dense wind.  We
demonstrate that the GISW model can produce a wide range of aspherical
flow patterns.  The dependence of the shock morphology on the initial
parameters conforms to the expectations of analytical models (Icke
1988).  We find that radiative cooling slows the evolution of the
forward shock by removing energy from the hot bubble and that
radiation heating and cooling changes the temperature structure of the
shocked slow wind material.

We have constructed self-consistent synthetic observations of the
models from forbidden line emissivities used in the energy loss term.
We present integrated intensity and long-slit spectrum,
(Position-Velocity) maps of the models projected at different angles
on the sky.  These synthetic observations are compared with real
intensity and Position-Velocity maps of PNe.  We find that there is a
very good match between the synthetic and real observations in terms
of morphologies, kinematics, and physical conditions.

 From the results of these simulations we conclude that the GISW scenario
can account for most, if not all, PNe morphologies, thus confirming Balick's
(1987) conjecture.

\vfill\eject

\section{Introduction}
Planetary Nebulae (PNe) take on a diverse range of shapes (Schwarz,
Corradi \& Melnick 1993).  In the last decade it has been recognized
that the morphology of PNe can be accounted for through the
interaction of concentric stellar winds. In this ``interacting stellar
winds'' (ISW) model of PNe, a fast tenuous wind from the PNe central
star expands into a slow, dense circumstellar envelope.  The envelope
was expelled as a wind from the progenitor AGB star (Kwok, Purton, \&
Fitzgerald 1979).  The ISW model has been successful in accounting for
many of the observational characteristics of spherical PNe
(Schmidt-Voigt\& K\"oppen 1986, Marten \& Sch\"onberner 1990,
Mellema~1993, Frank 1994 (Paper II), Mellema 1994 (Paper III)).  The
majority of Planetary Nebulae do not, however, show spherical
symmetry.  Instead, most PNe display elliptical or bipolar
morphologies (suggesting cylindrical symmetry).  Balick (1987)
suggested that the shapes of aspherical PNe could be explained through
a generalization of the ISW framework. In Balick's scenario the
progenitor AGB star expels its slow, dense wind with a toroidal
density distribution.  The fast wind from the PN central star then
drives an outflow pattern which is preferentially aligned with the
toroid's symmetry axis.  The theoretical consequences of this idea
were explored by Kahn \& West (1985).  By varying the slow wind's
equator to pole density ratio (e/p ratio), and considering the age of
the interacting wind system, Balick conjectured that the full range of
elliptical and bipolar morphologies could be recovered with the
generalized ISW (GISW) model.

Two-dimensional, analytical models of PNe (Icke 1988, Balick, Preston
\& Icke 1987, Icke, Preston \& Balick 1989) have successfully demonstrated that
elliptical and bipolar outflows can be produced by a GISW model.
Numerical simulations of the GISW scenario (Soker \& Livio 1989;
Mellema, Eulderink \& Icke 1991, Icke, Balick, \& Frank 1992a;
hereafter MEI91; IBF92) have recovered the analytical results and have
shown that the interaction of fast winds and circumstellar toroids
produces a rich variety of gasdynamical behavior.  In particular the
simulations of MEI91 and IBF92 have shown that the full nonlinear,
time-dependent solution of GISW gasdynamics produces flow patterns
with features not predicted in analytical models.  These features
include density enhancements along the poles, fish-tail shaped
features extending from the caps of the bipolar flow and well
collimated jets (Icke {\it et.\ al.} 1992b).  These flow patterns
allow the numerical models to embrace both ``classical'' elliptical
and bipolar nebulae (such as NGC 7662 and NGC 6905 respectively) as
well as PNe which do not fit into the elliptical or bipolar
classifications (e.g.\ the Red Rectangle and NGC 6369).

While the GISW scenario has had continued success in accounting for
the morphologies and kinematics of PNe, the majority of models to date
have treated the nebular gas without energy
sources or sinks. From observations, however, it is clear that PNe are
comprised of gas in different ionization states absorbing energy from
the central star and radiating it back into space. Radiation processes
can alter the gasdynamic evolution of PNe.  One dimensional
simulations have already demonstrated the significant role that
ionization by the stellar UV flux and cooling from forbidden lines
can play in the evolution of PNe.  Thus a proper treatment of
radiation-driven ionization, heating, and cooling is needed to fully
understand the dynamical evolution of PNe (Frank \& Mellema 1994,
henceforth Paper~I).  Radiation-gasdynamical
calculations also form the critical link between GISW models and
observations.  The calculation of observationally important forbidden
and permitted line volume emissivities should be intrinsic to any
radiation-gasdynamical model of PNe.  These emissivities, if
calculated individually as terms in the total radiative cooling, can
also be used to produce self-consistent ``synthetic observations'' of
the models.  These can be compared with real observations.

After having presented our methodology in Paper~I and two types of ISW
models in Papers II and III, this fourth paper contains the results of
numerical experiments in GISW radiation-gasdynamics.  The numerical
models we have constructed to perform these simulations incorporate
many of the radiation and microphysical processes \footnote \dag {We
distinguish between microphysical quantities and gasdynamic
macrophysical quantities.  Microphysical quantities are those not
directly relevant to the gasdynamics (which only feels the heating and
cooling).  We call optical depths, electron densities, ionization
fractions and gas temperature microphysical quantities.} thought to be
important in PNe. The results presented here comprise another step in
the numerical study of aspherical PNe evolution.  This paper is also
intended as a catalog of the generic intensity map morphologies and
kinematical patterns (in the optical) which can be obtained with the
radiation-gasdynamic GISW model.  We have already presented some
synthetic intensity maps in Frank {\it et. al.} (1993). In the next
paper in this series (Mellema \& Frank 1994, Paper~V) we will explore
parameter space more fully in an attempt to constrain models of AGB
wind asphericity as well as extend the study of synthetic observations
into other wavelength domains such as X-rays.

We note that the results presented in this paper should apply to the
formation of bipolar flows in many astrophysical settings.  Nebulae
surrounding symbiotic stars are thought to be close cousins of PNe and
have been modeled analytically with the GISW scenario by Henney \&
Dyson (1992).  Blondin \& Lundqvist (1993) have used GISW numerical
simulations to model the ring of SN1987A.  Their flow patterns are
similar to those of MEI91, IBF92 and those presented here. Another
example is the morphology of the Eta Carina nebulae (Meaburn,
Wolstencroft \& Walsh 1987).  It is quite similar to some bipolar PNe
and may have been produced through processes similar to those in the
GISW scenario. Bipolar molecular flows and Herbig-Haro jets may also
form through a variation of the GISW model (Frank \& Noriega-Crespo
1994).  Finally, at the largest scales relativistic jets like those
seen in AGN may be driven by GISW-type flows (Eulderink \& Mellema
1994).  We note that models of wind interactions with stratified
external medium have also been invoked to explain galactic
super-bubbles (MacLow, McCray, \& Norman 1989).

The organization of this paper is as follows. In section 2 of the
paper we discuss the numerical methodology and initial conditions used
in the simulations.  In section 3 we present the results of the models
in terms of their dynamical evolution.  In section 4 we present the
synthetic observations derived from the simulations and discuss their
relevance to observations of real PNe.  In section 5 we provide our
conclusions.

\section{Methodology and Initial Conditions}
The details and tests of the numerical method used to solve the
gasdynamical, radiation and microphysical equations has been described
elsewhere, (see references below for the gasdynamical methods and
Paper~1 for the radiation method).  We summarize the methodology as
follows: we solve the two-dimensional Euler (so no viscosity or heat
conduction are included), radiation transfer and ionization rate
equations in axisymmetric, spherical coordinates ($(r,\theta): v_\phi
= 0$).  The gasdynamical equations are solved using either a flux
corrected transport based scheme (the LCD method of Icke 1991) or a
Godunov based Roe-solver scheme (Eulderink 1993, Mellema 1993).  Both
gasdynamical codes are second-order accurate and can resolve shocks on
two or three grid-points. The radiation and ionization equations are
treated in parallel with the gasdynamics through operator splitting.

Radiation transfer from the central star is treated as a function of
wavelength.  We consider only the transfer of photons through H and
He.  Ionization was treated with a time-dependent method for H and He.
Ionization equilibrium is assumed for the heavier elements.  For
temperatures below $10^5$K energy losses through collisionally excited
and recombination line radiation is treated as cooling terms in the
energy density equation.  The forbidden, permitted, and recombination
line energy loss terms are further resolved into individual lines for
various elements (H,He,C,O,N,Ne) and ionization levels (I through V).
The emissivities for these lines was calculated using computationally
inexpensive fits to full 3 and 5-level calculations (Balick, Mellema
\& Frank 1993). At temperatures above $10^5$K we use the coronal
cooling rates of Dalgarno \& McCray (1972). We do not include the
heating and cooling of dust grains in the energy budget of the models.
While dust is expected to be important in young PNe (Zhang \& Kwok
1991), limited computational resources have forced us to neglect its
influence on the model.  We expect however that including dust will
produce more acceleration and more cooling during the early stages of
evolution.

The gasdynamic initial and boundary conditions for the models are the
same as those used in MEI91 and IBF92.  The fast wind is injected in
the first radial row of computational cells.  The slow wind
begins in the next radial row and continues out to the edge of the
grid.  The density in the slow wind is assumed to have the following
form, $$\rho(r,\theta) = {\rho_* \over F(\theta)}{\left({r_o \over
r}\right)}^2,\eqno(1)$$ $$F(\theta) = 1 - \alpha ({{e^{\beta\cos 2 \theta -
\beta} - 1} \over e^{-2\beta} -1 }). \eqno(2)$$
In the equations above $r_o$ is radius at which the slow wind begins.
The parameter $\alpha$ describes the pole to equator density contrast
(q) in the slow wind, e/p ratio $\equiv q = 1/(1-\alpha)$.  The
parameter $\beta$ controls the steepness of the pole-ward decrease in
density and, therefore, the shape of the slow wind.  Small values of
$\beta$, $(\beta \approx 1)$, give almost elliptical density contours
similar to those for a disk.  Large values of $\beta$, $(\beta \approx
6)$, yield density contours that are nearly spherical with a cone cut
out near the symmetry axis $(\theta = 0^\circ)$. As was discussed in
earlier papers the density distribution in equations (1) and (2)
represents a toroidal slow wind which was deposited with constant
velocity.  There is no special reason for using the distribution
represented by these equations.  We use it as a matter of definiteness
and, to some extend, tradition (Icke, Preston,
\& Balick 1989).  We note that in their study of SN1987A
Blondin \& Lundqvist (1993) used a number of density distributions as
initial conditions including an inverted form of equation (1) and (2).
The bipolar flow patterns they show in their results were
qualitatively similar to those presented here, in MEI91, and in IBF92.

The initial and boundary conditions for the radiation are as follows:
In all the models we present here the central star is assumed to be a
blackbody.  We chose stellar parameters (surface temperature, $T_*$,
and luminosity, $L_*$) to ensure that all the gas in the computational
grid would be ionized at least to the first stage. For this reason the
choice of stellar parameters may not be representative of the values
assumed during the entire lifetime of most PNe. They are, rather,
chosen to produce the basic ionization conditions observed in most
PNe. For the initial conditions we have chosen the stellar UV flux is
able to ionize oxygen up to O$^{2+}$.  The UV flux from the star is
continuously injected in the first row of computational cells and is
allowed to transfer radially through the grid.  We treat the diffuse
radiation field through the choice of appropriate recombination
coefficients (Osterbrock 1989).

We have run more than 70 simulations with different initial
conditions.  Of these we have selected five which demonstrate the
important generic features of the models' dynamical evolution and
observational appearance.  The initial conditions for these five
simulations are shown in table 1.  The initial conditions are
presented as mass loss rates ({\mdot}), and velocities (km s$^{-1}$)
for fast wind (fw) and slow wind (sw), the $\alpha$ and $\beta$ values
for the slow wind, and the parameters for the central star.  The
initial density in the slow wind, ($\rho_*$), in equation (1) can be
derived from the assumption of a constant wind velocity and
conservation of mass, $$\rho_* = {{\dot M} \over 4\pi v_{sw}
{r_o}^2}.\eqno(3)$$ All of the runs have the inner boundary set to
$r_o = 1\times 10^{16} cm$, a resolution of 100x100 cells on a quarter
of the computational plane ($\theta = 0-\pi/2$), and a radial cell
size of $\Delta r = 2\times10^{15} cm$.  Because of the detail
included in the radiation part of the numerical method the simulations
are computationally expensive.  We ended each simulation when the
forward shock ran off the grid.  Given our initial conditions this
comprised between 400 and 1200 years of simulated evolution and
required approximately 23 hours of Cray-XMP time.  We note that all
but one of simulations presented here were computed with the LCD
gasdynamical code.  Models computed with the Roe-solver gasdynamical
code produce nearly identical results (Mellema~1993).  We consider
this consistency to be an important confirmation of our methodology.

The models presented in this paper incorporate two significant
simplifications.  First, we consider the fast wind to begin
instantaneously after the (equally instantaneous) termination of the
slow wind.  Second, we hold the energy distribution of the stellar UV
flux constant throughout the simulation.  Thus our models do not
include the evolution of the inner boundary condition for either the
fast wind or the central star.  These are expected to be coupled
(Pauldrach {\it et. al.} 1988). It is clear that both the evolution of
the stellar UV flux and the coupled increase in fast wind velocity
will be important factors in determining the dynamics of PNe
(Papers~II--III; Mellema 1993).  In this study, however, we wish to
explore the radiation-gasdynamical effects of the GISW scenario in its
simplest form.  Since an investigation of this kind has not yet been
attempted, we believe that it is necessary to understand the
consequences of a simple model (with respect to boundary conditions,
not physical processes) first.  In later papers we will treat the more
complicated and uncertain details of time-dependent boundary
conditions and their effect on nebular evolution (Mellema 1993).
We note that we have not included the effects of heat conduction.
The presence of even a small, dynamically insignificant, magnetic
field will suppress heat conduction across field lines making its
influence of the PNe evolution negligible.  Heat conduction can be
observationally important for issues relating to X-ray emission
(Soker 1994).

We are also limited in these models to rather moderate spatial
resolution and short evolutionary times.
This is due to the large amounts of CPU times needed for the
simulations.  Since cooling due to radiation can produce significant
compression and collapse of the swept-up slow wind shell, the
restriction to moderate resolution constitutes a fundamental limitation
of these models. While our simulations clearly resolve the morphologies
of the fundamental gasdynamic discontinuities (e.g.\  shocks) some
details of flow patterns such as instabilities do not appear in these
models.  We do not contend that we have the resolution needed such that
the simulations have converged. We are currently modifying the code to
reduce some of the radiation detail and allow faster run-times with
higher resolution.  However, in the present study we are interested
primarily in the evolution and observational appearance of global
morphology and kinematic patterns in PNe.  Thus the lack of higher
resolution does not present a challenge to the results and conclusions
presented here.  We note that high resolution GISW simulations with
cooling performed by Blondin and Lundqvist did not show any gross
changes in morphology from their lower resolution non-radiative runs.

\section{Results: Gasdynamical Flow Patterns}
The basic gasdynamical flow pattern associated with the ISW scenario
consists of an interaction region formed by a triplet of
discontinuities.  The interaction region is bounded on the inside and
outside by the undisturbed stellar winds. (Weaver {\it et. al.} 1977;
Dyson \& Williams 1980).  Moving radially outward from the star one
first encounters an inner shock where the fast wind (fw) is
decelerated, compressed and thermalized. This thermalization creates a
subsonic, high temperature region downwind of the inner shock.  For
the fast winds considered in the PNe problem the pre- and post-shock
densities are rather low.  We refer to the region of high temperature
shocked fast wind as the hot bubble (hb).  The low densities and high
temperatures in the hot bubble imply cooling times longer than the
evolutionary time-scale for of most PNe.  Thus the hot bubble will not
be able to cool significantly during the PN's lifetime.  At the outer
edge of the interaction region is a forward shock. This shock wave is
driven into the undisturbed slow wind by the thermal pressure of the
hot bubble.  Slow wind (sw) material which passes through the forward
shock is compressed and heated forming a dense shell (ds).  Unlike the
material in the hot bubble, the cooling time-scale for the dense shell
is short compared to the evolutionary time-scale of the PN.  Thus when
radiation losses are included the shocked slow wind material quickly
radiates away its newly acquired thermal energy.  The radiation losses
make the forward shock isothermal, ($T_{ds} = T_{sw}$), and produces
further compression in the shell.  Separating the dense shell of
shocked, swept-up slow wind material and the hot bubble of shocked
fast wind material is the third gasdynamic discontinuity which is a
contact discontinuity (cd).  Material can not diffuse across the cd
and mixing can only occur via two or three-dimensional motions
(e.g. instabilities, turbulence, etc.)

With this sketch in mind we present in Fig 1 greyscale representations
of the logarithm of gas density for the four models A, B, C and D.
The models are shown at six different times in their evolution. Notice
first that the basic triple discontinuity is apparent in each of the
models.  Consider the final evolutionary frame for each of the
models.  The forward shock and dense shell appear as the darkest
greyscales.  Typical densities are $\langle n_{ds}\rangle \approx 10^3
cm^{-3}$, and $\langle n_{hb}\rangle \approx 1 cm^{-3}$. Interior to
the inner shock is the freely expanding undisturbed fast wind.  In
this region the density decreases from its value $\rho_{fw}$ at the
inner radius of the grid becoming geometrically diluted as it expands
($\rho \propto \rho_{fw} (r_o/r)^{-2}$).

Fig 1 demonstrates the dependence of model morphology on key initial
parameters.  In his analytical study of GISW morphologies Icke (1988)
derived an expression for the evolution of the forward shock, $r_s$,
using Kompaneets (1960) formalism, $$r_s = r_s(t,\theta),\eqno(4a)$$
$${\partial r_s \over \partial t} = {\lbrace A (1+({1 \over
r_s}{\partial r_s \over \partial \theta} ))\rbrace}^{1
\over 2},\eqno(4b)$$ $$A = {\gamma + 1 \over 2} {P_{hb} \over
\rho_{sw} (\theta)},\eqno(4c)$$ Icke demonstrated that the evolution
of the forward shock was controlled by the ``acceleration parameter''
A which is the local ratio of driving force for the shock to the
external inertia ($P_{hb}$ is the pressure in the hot bubble). In our
study we confirm the expectations of Icke's model. The e/p ratio q,
the initial shape of the slow wind ($\beta$), and the fast wind
kinetic energy density $e_k = \rho v_{fw}^2$ determine the dynamical
shape of both the inner and outer shocks. These are the parameters
which also determine Icke's acceleration parameter A.  The e/p ratio
(through $\alpha$) and the parameter $\beta$ are needed to determine
the initial slow wind density distribution F($\theta$) (equations 1
and 2).  In addition $P_{hb}$ is equivalent to the fast wind energy
density $e_k$.  We now consider the role of these input parameters on
the shock morphology seen in the numerical simulations.

We examine first the role of the pole to equator density
contrast. Model A with $q = 2$ maintains an elliptical morphology
throughout its evolution.  Models B, C and D, with $q = $ 3.33, 5 and
3.33 respectively, quickly assume bipolar morphologies which become
more pronounced as the models evolve.  Here we define bipolar
morphology as a forward shock configuration with one or more cusps
separating an equatorial belt (a waist) from the bipolar lobes.  From
analytical models it can be shown that the cusps represent the
intersection of the shock which forms the bipolar lobes with the shock
that makes up the equatorial ``barrel''.  We find that $q > 2$
produces bipolar morphologies while $q < 2$ produces elliptical shock
configurations.  This may be an important point for theories which use
common envelope evolution in binary stars (Soker \& Livio 1989; Soker
1993; Livio 1993) to explain the origin of the slow wind geometry.
The shape of the inner shock depends on the e/p ratio.  Note that for
$q < 2$ the inner shock is almost spherical while for $q > 2$ it
assumes a more aspherical, prolate geometry.

Varying the fast wind energy density, $e_k$, changes both the
morphology and speed of the models' evolution. Models B and D have the
same e/p ratio and $\beta$.  Model D however has a kinetic energy
density that is approximately three times larger than that in Model B.
Comparison of their evolution in fig 1 shows that Model D develops a
narrower waist band than Model B. Model D also evolves faster,
expanding off the grid ($r = 2\times 10^{17} cm$) in 400 y compared
with 800 y for model B.  Note also that the inner shock is dependent
on the fast wind energy density.  The inner shock in Model D assumes
the most prolate configuration of the models presented here.

The initial shape of the slow wind, determined by $\beta$, determines
the details of the forward shock morphology such as where the cusps
appear.  We will see in the next section how this affects the
observational appearance of the models.

We note that the morphologies which develop in the
radiation-gasdynamic simulations presented here do not differ
qualitatively from those produced by the pure gasdynamic simulations
of MEI91 or IBF92.  We do find that the radiative models evolve with
lower forward shock velocities as was found in one-dimensional
spherically symmetric models (Paper~I).  The velocity of
the contact discontinuity, which acts as the piston driving the
forward shock, is determined by the pressure in the hot bubble.  Thus
the difference in the shock velocities between the non-radiative and
radiative models can be attributed to the energy lost from the hot
bubble.  The energy losses in the hot bubble come, primarily, from
free-free emission.  In their GISW simulations Blondin \& Lundqvist
found a similar result between their radiative and non-radiative
models.

In order to complete the description of the dynamical evolution we
present in figs 2 and 3 velocity and temperature maps taken from the
models.  Fig 2 shows the velocity field in Model A and D before the
forward shock is driven off the grid.  Only those velocity vectors
with $|v| < |v_{fw}|$ are plotted.  The inner, first row of vectors
traces out the inner shock.  In model A the inner shock is nearly
spherical with only a mild prolate deformation.  As was observed in
the non-radiative simulations, the weak prolate geometry of the inner
shock produces spatial oscillations in the post-shock, hot bubble
velocity.  The origin of this effect is primarily numerical, occurring
when the inner shock jumps from one radial row of cells to the next.
The inner shock does however appear to be subject to a corrugation
instability (Landau \& Lifshitz 1987) as even high resolution
non-radiative runs show a periodic rippling in the post-inner shock
velocity.  We conjecture that acoustic noise in the hot bubble,
perhaps produced by a ringing mode of the cavity (Frank \&
Noriega-Crespo 1994), drives local oscillations in the shock normal.
These oscillations then cause variations in deceleration through the
shock.  We note, however, that at the present time the nature of these
oscillations remain uncertain and will be the subject of a next paper
(Paper~V).

Another velocity effect which appears in both the radiative and
non-radiative models is flow focusing due to the asphericity of the
inner shock.  This flow focusing was anticipated by Icke (1988) and
was first explored numerically in the GISW models performed by Soker
\& Livio (1989).  It was later investigated more fully in the
higher-order accurate models of MEI91 and IBF92.  The velocity field
for Model D is shown in the right panel of fig 2.  It shows the
variation of the shock normal relative to the radially streaming fast
wind and the effect this has on the post-shock velocity field.  When
the inner shock is highly prolate the fast wind encounters the shock
at an oblique angle.  Only the normal component of the fast wind,
relative to the shock, will be affected upon passage through the
shock.  Thus the downstream velocity vector of a parcel of gas will be
both shorted and refracted relative to the upstream vector.  Given its
prolate geometry the inner shock in Model D acts as a lens focusing
the post shock flow towards the symmetry axis.

In fig 3 we present a surface plot of the temperature distribution in
Model C after 783 years of evolution.  The central hole in the
temperature distribution marks the freely expanding fast wind region
where temperatures are low ($T\sim 10^4$ K).  The "cliff walls" mark the
location of the inner shock. The hot bubble dominates the surface plot
appearing as the wide peanut-shaped plateau of high temperature.  We
can estimate the temperature that should be achieved in the hot bubble
by using the jump conditions for a strong shock $$P_2 = ({2 \over
{\gamma - 1}})\rho_1 {v_s}^2,\eqno(5)$$ $$\rho_2 = ({{\gamma + 1}
\over {\gamma - 1}})\rho_1,\eqno(6)$$ where the subscripts 1 and 2 refer
to the preshock and post-shock states and $v_s$ is the shock velocity.
Observe from fig 1 that the inner shock is slowly expanding.  Thus in
the shock frame $v_s \sim v_{fw}$ Combining this fact with equations
5, 6 and the equation of state we conclude that $$T = {2\over{\gamma
+1}^2}{m \over k} {v_s}^2 \sim 3\times 10^7 K.\eqno(7)$$ Fig 3
demonstrates that the temperatures in the hot bubble achieved in the
simulations are of this order.  Detailed inspection of the temperature
distribution in the hot bubble shows, however, that the actual
temperatures are slightly lower, an effect which can be ascribed to
the slow expansion of the inner shock and the radiative losses in the
hot bubble.

Fig 3 also demonstrates the effect of the stellar UV flux on both the
shocked and undisturbed slow wind gas.  In our initial conditions
$T_{sw} = 200 K$ which is appropriate for a cool AGB wind.  The UV photons
streaming away from the star have ionized and heated the undisturbed
slow wind.  The ionization also introduces strong collisionally excited
forbidden and permitted line cooling into the energy equation.  A new
thermal balance is obtained at $T_{sw} \sim 9000$ K throughout the
undisturbed slow wind.  The shocked slow wind in the dense shell
also cools effectively.  Again the cooling is due to collisionally
excited lines, in this case radiating away the thermal energy
gained in passing through the forward shock. Fig 3 shows that the
forward shock is nearly isothermal.  The temperatures in the shell are
at most a few thousand degrees higher than that in the undisturbed
slow wind.  Note that the highest shell temperatures occur at the
polar caps.  This is to be expected as the densities there are lower
and the cooling $L$, which goes as $L \propto \rho^2$, is less effective.

\section{Results: Synthetic Observations}
In this section we present projected intensity and Position-Velocity
(long slit spectra) maps derived from the models.  Using the volume
emissivities calculated for the total radiative cooling, $L$, we
produce self-consistent synthetic observations of the models.  In
this section we also discuss the comparison of these synthetic
observations with real observations of PNe.

The projected intensity and Position-Velocity (P-V) maps are
calculated from the models as follows (see also Henney \& Dyson 1992):
The two-dimensional volume emissivity $\epsilon(r,\theta)$ from a
particular forbidden, permitted, or recombination line is rotated
about the symmetry axis to produce a three-dimensional emissivity
distribution.  This distribution is then mapped on to a Cartesian grid
$(x,y,z)$ where the $(x,y)$ plane is considered to be the plane of the
sky. The long axis of the model nebulae is placed along the y-axis.
The three-dimensional emissivity $\epsilon(x,y,z)$ was then tilted at
some angle $i$ with respect to the y axis and then integrated along
the line of sight to form the projected intensity map $I(x,y;i)$.  We
will refer to $i$ as the inclination angle of the projection.  The P-V
diagrams were constructed in a similar manner where the tilted
emissivity is first binned according to the projected velocity along
the line of sight $v_l$ at each point.  The P-V diagrams are
constructed with the slit set across the long axis of the nebulae $(y
= 0)$. The true expansion velocity and $v_l$ are then related through,
$v_l = v_{exp} \sin i$. The binned emissivities are integrated to give
the long-slit spectrum P-V map $S(x,v_l;i)$.

In figs 4, 6, 8 and 10 we present $I(x,y,i)$ for the models. In figs
5, 7, 9 and 11 we present $S(x,v_l;i)$ for the models. The projected
intensities and P-V diagrams are calculated using the volume
emissivity of the [OIII]$\lambda$5007 line and are shown at six
different inclination angles $i = 12^\circ, 25^\circ, 45^\circ,
60^\circ, 80^\circ$ and $90^\circ$. We consider each of the models and
its comparison with observations separately.  In what follows below we
refer to the objects appearing in the synthetic observations as model
nebulae.

Fig 4 demonstrates, for model A, the effect of local variations of
$\epsilon([OIII])$ and projection on the intensity maps.  The
brightest parts of the nebulae appear in the dense shell along the
equator where $\epsilon([OIII]) \propto \rho^2$ is highest.  The
nebulae fades near the symmetry axis due to the decreasing density
near the poles.  At low projection angles the morphology of the model
nebulae mirrors that of the density distribution and is essential
elliptical.  An excellent match to this morphology is seen in M 3-9
(Schwarz, Corradi \& Melnick 1993). For large $i$, when we look down
the pole of the model, the morphology is circular.  Note that for $i =
25^\circ$ emission from the front and back of shell intersect forming an
hourglass shape across the middle of the model nebulae.  This
morphology is similar to that seen in the Dumbbell Nebula (NGC 6853).
The P-V diagrams for Model A shown in fig 5 indicate that the
expansion velocity for equatorial regions of the shell is $v_{exp}
\sim 55 km s^{-1}$.  This can be seen from the P-V map at low $i$.  In
the $i=90^\circ$ P-V map the velocity of the shell at the poles is
seen to be higher ($v_{exp} \sim 95 km s^{-1}$).  The skewing of the
figure as $i$ increases is to be expected (O'Dell \& Ball 1985). As
the shell is tilted, the higher velocities along the poles are
projected onto the line of sight in blue- and red-shifted components
which skew the kinematic figure.  Note that the emission from the slow
wind material is seen in the P-V maps both inside and outside the
skewed elliptical shell.  The slow wind appears skewed up until
$i=90^\circ$. NGC 7662 shows a kinematical pattern (Chu 1989) that is
qualitatively similar to that seen in the P-V maps at low $i$ for
model A though no slow wind emission internal to the shell appears.

For Model B, fig 6 demonstrates that the cusps in the shell at
mid-latitudes (coordinate $\theta$) strongly modify the projected
intensity maps. Recall that the cusps form where lobe and equatorial
shocks meet.  The intersection of these shocks produces enhanced
densities in the shell.  The density enhancements appear as rings in
the projected intensity maps.  At low $i$ the model nebulae appear as
limb-brightened barrels with fainter lobes protruding from the
caps. As $i$ increases the rings overlap producing the appearance of
eyes in the model nebula.  For $i \ge 60^\circ$ the rings are almost
co-extensive and the eyes disappear. With increasing $i$ an ellipse
interior to the limb-brightened circular region appears.  These patterns
are observed in a number of PNe. The general agreement between a
simulation with similar initial conditions as Model B and observations
of PNe morphologies was discussed in Frank {\it et. al.} (1993) (see
their fig 2).  They found that the morphologies of the projected
intensity maps shown in fig 6 matched well with (in order of
increasing $i$) NGC 40, NGC 7354, NGC 3587 (the Owl), Abell 82, NGC
2392 (the Eskimo) and NGC 1535.  The P-V maps of Model B presented in
fig 7 show that the lobes are expanding at high velocities $v_{exp}
\sim 100 km s^{-1}$ In the equatorial regions $v_{exp} \sim 50 km
s^{-1}$.  Note that as the inclination becomes very high, $i >
60^\circ$, the kinematical pattern opens up and the dense shell is not
continuous.  This effect occurs when the the line of sight cuts
through the throat of a bipolar nebulae exposing high velocity, low
emissivity gas.  Exactly such a pattern is observed in the kinematical
studies of the Eskimo Nebula, NGC 2392 (O'Dell \& Ball 1985, O'Dell,
Weiner \& Chu 1990).  In those papers the bright shell in the observed
P-V map was open with faint, high velocity ``plumes'' at positions
interior to the shell.  The velocity of the shell and plume were
calculated to be $v = 60$ and $v \approx 125 \ km \ s^{-1}$
respectively.  The inclination angle estimated for the Eskimo was $i
\sim 70^\circ$.  These shell and plume velocities in model B values
are in good agreement with these values.

The intensity maps presented in fig 8 show that Model C maintains the
appearance of a cylinder up to high inclinations. The overlap of the
bright regions near the equator produce elliptical patterns interior
to the ``walls'' of the cylinder.  The morphologies shown in fig 8 are
consistent with those seen in A55, IC 4406, He2-36, He 2-64, He
2-141. Mz 1, and Na 2 (Schwarz, Corradi \& Melnick 1993). The P-V maps
for Model C presented in fig 9 show the higher velocities of the lobes
relative to the equatorial regions.  This can be seen in the hourglass
shape of the shell.  As the projection angle increases the shell
assumes a distribution which is more like a skewed rectangle.  In
their study of He 2-36 Corradi \& Schwarz (1993a) observed a P-V
distribution (see their figure 2) similar to that seen in fig 9 for
$i=45^\circ$.  In their analysis they concluded that velocities in the
nebula range from $45$ to $80 \ km s^{-1}$ with a projection angle of
$i = 30^\circ$.  They also found that the number densities in the
nebulae at the poles are $n = 600 cm^{-3}$.  The morphologies shown in
the Corradi \& Schwarz maps as well as the physical conditions they
derive from their observations are in good qualitative agreement with
our model C and synthetic observations derived from it.  We note
however that the appearance of point symmetry in the brightness
distribution of the lobes in He 2-36 can not explained by our models.
We will come back to this point in the last section.

The intensity maps presented in fig 10 show that, at low inclinations,
($i < 45^\circ$), Model D assumes a double horned shape .  At higher
$i$ the projected intensity assumes the cusp-dominated double ring
``eye'' appearance seen in Model B.  Recall that model D has a high
fast wind kinetic energy density.  The strength of the wind-wind
interaction in this model drives the less dense polar lobes to a high
velocity leaving the shock in the equatorial region behind at smaller
radii.  Thus the emissivity in the dense, low latitude, small radii
regions dominates over that at the poles. Note that there is
significant emission above the cusps which was not seen in model B.
Likewise the rings which form the eyes are brighter in Model D than in
Model B.  This double-horned/bright double-ringed configuration has
been observed in the symbiotic stars He 2-104, BI Cru and the
planetary nebula MyCn18 by Corradi \& Schwarz (1993b) who refer to it
as a Crab morphology.  The opening angle of the horns can also be
thought of as the ratio of the width of the dense shell at the equator
to that at middle of the lobes In He 2-104 and BI Cru the opening
angle is higher than in fig 10.  We attribute this to differences in
age.  Our model has evolved for only $t = 330 y$ while Corradi \&
Schwarz estimate the age of He 2-104 and BI Cru to be $t\sim 10^3 y$.
The P-V maps for Model D are shown in fig 11. The high expansion
velocities in the lobes are apparent at low inclinations. At high $i$
very high velocities show up in the interior of the cavity (as was
already discussed for Model B).  It is important to note that at low
$i$ the P-V maps show the lobes flaring out to higher velocities than
the equatorial region.  By connecting the patterns for the lobes we
reproduce the skewed {\bf X} configuration observed for He 2-104
(Corradi \& Schwarz 1993b, figure 4). Note that unlike our [OIII] P-V
map, their kinematic observations were taken in the light of a low
ionization [NII] forbidden line which picks up emission at the poles
much like the anomalous FLIERS discussed by Balick et al.\ (1993ab).
These low ionization knots are not seen in our simulations.  The
kinematic pattern displayed by the double ring nebulae MyCn18 (Corradi
\& Schwarz 1993b, figure 10) shows a skewed column of three ellipses
similar to our $i = 45^\circ$ P-V maps for Model D. We note however
that a better match seems to be found in Model B at a similar
inclination angle.

Finally we present a side by side comparison of real observations and
a model. Fig ~12 shows the observed [OIII]$\lambda5007$ image and
major axis long slit spectrum for NGC~3242 (data from Balick~(1987)
and Chu~\& Jacoby (private communication) respectively), next to the
corresponding synthesized data taken from Model~E. Model~E was
calculated using the Roe~solver gasdynamic scheme (see Sect.~2). It
uses a high value for $\beta$ (7.0), which leads to the formation
small polar extrusions.  Similar extrusions are seen in NGC~3242. We
estimate the inclination of this object to be in the range
$20^\circ$~-- $30^\circ$. Therefore the synthesized data are displayed
at $i=25^\circ$.Both the observed and the synthesized P-V~diagram have
a skewed elliptical shape with a characteristic point-symmetry. Note
how well the synthesized P-V~diagram matches the position of the
brightest parts of the spectrum. These correspond to the rings formed
by the cusps.  NGC~3242 is surrounded by a diffuse outer envelope
which shows a higher expansion velocity than the inner bright
rim. This is not found in our simplified model. However, simulations
in which an evolving star star is used do show the formation of such
envelopes (Mellema~1993; Paper~III).  We note that the work of Soker,
Zucker, \& Balick (1992) provides good data on the initial
configuration in the slow wind of NGC 3242.  In principle this data
would be better to use as an initial configuration for a model of NGC
3242.  However our purpose in the present paper is to show that the
use of a single family of initial conditions can reproduce many of the
observed features in real PNe.  The use of more realistic initial
conditions for detailed comparisons with individual PN will be a
future project.

In closing we must note a major deficiency of the models: their
inability to produce the fast moving, low ionization knots (called
ansae or FLIERS) aligned along the major axis of the nebulae (Balick
{\it el at} 1993ab).  Thus, as discussed in Soker (1990) and verified
in MEI91 and IBF92, GISW gasdynamics does not appear able to produce
ansae in a natural way.  The production of ansae remains a mystery
which may require the addition of special evolutionary conditions at
the early stages of PN formation.

\section{Summary and Conclusions}
5.1 $\underline {Summary}$: We have presented radiation-gasdynamic
simulations of aspherical Planetary Nebula (PN) evolution.  These
simulations were constructed using the Generalized Interacting Stellar
Winds scenario where a fast, tenuous outflow from the central star
expands into a toroidal, slow, dense circumstellar envelope.  We have
demonstrated that the GISW model can produce aspherical flow patterns.
In particular we have shown that by varying key initial parameters we
can produce a variety of elliptical and bipolar forward shock
configurations.  The dependence of the shock morphology on the initial
parameters conforms to the expectations of analytical models (Icke
1988).  We have demonstrated that including radiation-transfer,
ionization, and radiative heating and cooling does not drastically
alter the global morphologies.  Radiative cooling does slow the
evolution of the forward shock by removing energy from the hot bubble.
The evolution of the forward shock configuration is independent of the
ionization of the undisturbed slow wind.  Also, radiation heating and
cooling does change the temperature structure of the shocked slow wind
material compressed into the dense shell.  As is expected for
radiative shocks we find $T_{ds}
\approx T_{sw}$.

We have also constructed self-consistent synthetic observations of
five typical models from forbidden line emissivities used in the
energy loss term.  We presented integrated intensity maps and
long-slit spectrum, (Position-Velocity), maps of the models projected
at different angles on the sky.  These synthetic observations were
compared with real intensity and P-V maps of PNe.  This paper is
intended as a survey of radiation-gasdynamic GISW simulation results.
Thus in these comparisons we have looked for qualitative agreement
between models and reality in terms of morphological and kinematic
patterns as well as actual values of physical parameters such as
velocities and densities.  We find that there is a good match between
the synthetic and real observations.

5.2 $\underline {Conclusions}$: Based on the results presented above
we conclude that the GISW scenario proposed by Balick (1987) can
explain the majority of PNe morphologies and kinematics. While our
models contain a number of simplifications with respect to the
evolution of the central star and the fast wind it seems clear that
the elliptical, barrel-shaped, bipolar, double-ring, and owl-eyed
patterns seen in many aspherical PNe can be accounted for quite easily
through the radiation-gasdynamic version of the GISW model (with
projection effects included).  The same conclusion holds for the
kinematic patterns.  The correspondence of the observed P-V map
patterns for the Eskimo nebula with the high inclination-angle
synthetic observations of models B and D is quite good.  This in spite
of the fact that the initial conditions for those models were not
picked with any particular real PNe in mind. Thus while there still
remains a number of important questions to be resolved, we feel that
the issue of global PNe morphologies has been resolved by the GISW
model.  { \bf Interacting winds can account for most, if not all, PNe
morphologies}.  Thus we confirm Balick's original suggestion and the
results of earlier non-radiative numerical studies (Soker \& Livio
1989, MEI91, IBF92).  While it is still possible that other process
(such as magnetic fields, Chevalier \& Luo 1993) may play a role in
shaping PNe it is clear that they are not {\it needed} to shape PNe.

With respect to PNe morphologies three issues stand out which need to
be resolved.  First, the production of ansae remains an enigma. The
solution to this problem appears to require a better understanding of
the late stages of AGB star evolution (including binaries and common
envelopes, Soker 1993).  Second, as was noted before, the question of
what effect the evolution of the central star and fast wind have on
the evolution of the nebula.  To answer this question one us is
currently undertaking computations with time-dependent inner boundary
conditions (Mellema~1993).  The third question which needs to be
addressed is the striking point symmetry which often occurs in PNe
(Schwarz, Corradi, \& Melnick 1993).  It has been conjectured that
point symmetry occurs due to precession of the symmetry axis of the
nebulae, something which may be tied to the binary nature of the
central source. All of our models are, by assumption, axisymmetric.
The synthetic observations, therefore, produce plane or mirror
symmetric model nebulae. Any departure from axisymmetric geometry
requires three-dimensional models which can possibly be numerically
simulated with the latest generation of supercomputers. These
three-dimensional models will, however, probably need to be
non-radiative and low resolution.

Finally, if one accepts our conclusions then the most important
question to be asked is not about PNe but about AGB stars: What
mechanism produces the dense, toroidal wind at the end of the stars
evolution up the AGB?  A number of possible mechanisms have been
proposed (e.g. rotation, magnetic fields). However, the most promising
appears to be gravitational interaction of binary stars either through
common envelope evolution or tidal friction (Livio 1993; Soker 1993).
One of the interesting possibilities that emerges from this paper is
the use of PNe morphologies as a diagnostic for theories of AGB wind
shaping, in particular the binary star hypothesis.  If the GISW model
continues to have success in finding accord with observations of PNe
then a study of initial slow (AGB) wind parameters, their relation to
model PNe morphologies, and the statistics of real PNe morphologies
(Stanghellini, Corradi \& Schwarz 1993) may provide constraints for
theories of AGB wind shaping.  In this way the ``What'': aspherical
PNe, and the ``How'': GISW radiation-gasdynamics might be connected to
the ``Why'': AGB wind shaping.

{\bf Acknowlegements} We wish to thank our advisors Bruce Balick and
Vincent Icke for the guidance throughout the many years of this
project. R. Corradi provided the what, why and how
formulation. You-Hua Chu kindly let us use some of her
observations. A.F. wishes to thank Alberto Noriega-Crespo for his help
and Leiden University for its' endless hospitality. A.F. enjoyed
support from NSF grants AST 89-13639, INT-9200916, AST 91-00486, and
the Minnesota Supercomputer Institute. G.M.'s work on PNe was funded
by NWO(ASTRON), grant no.\ 782-372-029.
\bigskip
\bigskip
\vfill\eject

\centerline{\bf Bibliography}

\def\ref{\hang\noindent}

\ref Balick, B., 1987, \aj 94, 671

\ref Balick, B., Preston, H.L., Icke, V., 1987, \aj 94, 1641

\ref Balick, B., Mellema, G., Frank, A. 1993, \aa 275, 588

\ref Balick, B., Rugers, M., Terzian, Y., Chengalur, J.N., 1993a, \apj 411, 778

\ref Balick, B., Perinotto, M., Maccioni, A., Terzian, Y., Hajian, A.,
1993b, submitted to ApJ

\ref Blondin, J.M., Lundqvist, P., 1993, \apj 405, 337

\ref Chevalier, R.A., Luo,D., 1993, preprint

\ref Chu, Y. 1989, in: IAU Symposium 131: Planetary Nebulae, S.
Torres-Peimbert (ed.). Reidel, Dordrecht, p.~105

\ref Corradi, R.L.M., Schwarz, H.E., 1993a, \aa 268, 714

\ref Corradi, R.L.M., Schwarz, H.E., 1993b, \aa 269, 462

\ref Dalgarno, A., McCray, R. 1972, ARA\&A 10, 375

\ref Dyson, J.E., Williams, D.A., 1980, Physics of the interstellar
medium, Wiley, New York

\ref Eulderink, F., 1993, PhD thesis: Numerical relativistic
hydrodynamics, University of Leiden

\ref Eulderink, F., Mellema, G., 1994, submitted to A\&A

\ref Frank, A., Balick, B., Icke, V., Mellema, G., 1993, \apj 404, L25

\ref Frank, A., 1993, \aj, in press (Paper II)

\ref Frank, A., Mellema, G. 1994, \aa, in press (Paper I)

\ref Frank, A., Noriega-Crespo, A. 1994, \aa, in press

\ref Henney, W.J, Dyson, J.E., 1992, \aa 261, 301

\ref Icke, V., 1988, \aa 202, 177

\ref Icke, V. 1991, \aa 251, 369

\ref Icke, V., Preston, H.L., Balick, B., 1988, \aj 97, 462

\ref Icke, V., Balick, B., Frank, A. 1992a, \aa 253, 224 (IBF92)

\ref Icke, V., Mellema, G., Balick B., Eulderink, F., Frank, A. 1992b,
\nat 355, 524

\ref Kahn, F.D., West, K.A., 1985, \mn 212, 837

\ref Kompaneets, A.S., 1960, Dokl. Akad. Nauk SSSR 130, 1001

\ref Kwok, S., Purton, C.R., Fitzgerald, P.M., 1978, \apj 219, L125

\ref Landau, L.D., Lifshitz, E.M., 1987, in: Course of Theoretical
Physics VI, Fluid Dynamics, Pergamon Press, Oxford

\ref Livio, M., 1993, in: IAU Symposium 155: Planetary Nebulae,
R.~Weinberger, A.~Acker (eds.). Kluwer, Dordrecht, p.~279

\ref Livio, M., Soker, N., 1988, \apj 329, 764

\ref MacLow, M.M., McCray, R., Norman, M.L., 1989, \apj 337, 141

\ref Marten, H., Sch\"onberner, D. 1991, \aa 248, 590

\ref Meaburn, J., Wolstencroft, R.D., Walsh, J.R., \aa 181, 333

\ref Mellema, G., 1993, PhD Thesis: Numerical models for the formation
of aspherical planetary nebulae, Leiden University

\ref Mellema, G., 1994, submitted to \aa (Paper III)

\ref Mellema, G., Eulderink, F., Icke, V. 1991 \aa 252, 718 (MEI91)

\ref Mellema, G., Eulderink, F., 1994 \aa, in press

\ref O'Dell, C.R., Ball, M.E., 1985, \apj 289, 526

\ref O'Dell, C.R., Weiner, L., Chu, Y.-H., 1990, \apj 362 226

\ref Osterbrock, D.E. 1989, Astrophysics of Gaseous Nebulae and Active
Galactic Nebulae, Oxford University Press

\ref Pauldrach, A., Puls, J., Kudritzki, R.P., M\'endez, R.H. Heap,
S.R., 1988, \aa 207, 123

\ref Schmidt-Voigt M., K\"oppen, J., 1987, \aa 174, 211

\ref Schwarz, H.E., Corradi, R.L.M., Melnick, J. 1992, A\&A Suppl.\ 96, 23

\ref Soker, N., 1990, \aj 99, 1896

\ref Soker, N., 1992, \aj 104, 2151

\ref Soker, N., 1993, in: Mass Loss on the AGB and Beyond,
H.E.~Schwarz (ed.). ESO, Garching bei Munchen, p.~18

\ref Soker, N., Livio, M. 1989, \apj 339, 268

\ref Soker, N., Zucker, D.B., Balick, B., 1992 \aj 104, 2151

\ref Soker, N., 1994, \aj 107, 261

\ref Stanghellini, L., Corradi, R.L.M., Schwarz, H.E., 1993, \aa 279, 521

\ref Weaver, J., McCray, R., Castor, J., Shapiro, P., Moore, 1977,
\apj 218, 377

\ref Zhang, G., Kwok, S., 1991, \aa 250, 179

\bigskip
\bigskip
\vfill\eject

\centerline{Figure Captions}
Fig 1. Evolution of the density for models A,B,C and D.  Greyscale
representations of log$_{10}$ are shown at six different times for
each model.  The final frame for models A,B and C represent t = 783 y
of evolution.  The final frame for model D represents t = 350 y of
evolution.  In the final frame for all the models the darkest
greyscale represents $\log_{10} (\rho \ [g cm^{-3}])$ = -20.19.  The
lightest greyscales represent $\log_{10} (\rho \ [g cm^{-3}])$ =
-23.88.  The linear scale of each figure is $4\times 10^{17}$ cm
square. Note the shape of the inner shock in the different models.

Fig 2. Velocity vectors plotted for models A and D after t= 783 and 350
y respectively.  For model A all velocities less than 1000 km
s$^{-1}$ are plotted.  For model D only velocities less than 2500
km s$^{-1}$ are plotted.  The x and y axes are plotted in units of
$1\times 10^{16}$ cm. Note the corrugation of the asphericity of the inner
shock its effect on the flow there.

Fig 3.  Surface plot of log$_{10}$ (temperature) for model C after t = 783
y.  The x and y axes are plotted in units of $1\times 10^{16}$ cm. Note
the high temperatures in the hot bubble and the nearly constant
temperatures in the undisturbed slow wind and swept-up shell.

Fig 4. Projected [OIII]\lam 5007 intensity map for model A taken after
t = 783 y of evolution. The x and y axes are plotted in units of
$1\times 10^{16}$ cm.  The projection angle is, from left to right and
top to bottom, i = 12, 25, 45, 60, 80, 90$^\circ$. See text for
details.

Fig 5. Projected [OIII]\lam 5007 long slit spectrum Position-Velocity
map for model A taken after t = 783 y of evolution. The position axis
is plotted in units of $1\times10^{16}$ cm. The velocity axis is
plotted in units of km s$^{-1}$ The projection angle is, from left to
right and top to bottom, i = 12, 25, 45, 60, 80, 90$^\circ$. See text
for details.

Fig 6.  Same as Fig 4 for model B.

Fig 7.  Same as Fig 5 for model B.

Fig 8.  Same as Fig 4 for model C.

Fig 9.  Same as Fig 5 for model C.

Fig 10.  Same as Fig 4 for model D. Figure taken at t = 350 y of evolution.

Fig 11.  Same as Fig 5 for model D. Figure taken at t = 350 y of evolution.

Fig.~12. Comparison of real and synthesized data. Top left:
[OIII]$\lambda5007$ image of NGC~3242 (from Balick~1987). Bottom left:
[OIII]$\lambda5007$ long slit echelle spectrum along the major axis of
NGC~3242 (from Chu~\&\ Jacoby, private communication). Top right:
synthesized [OIII]$\lambda5007$ intensity map for model~E, taken after
$t=1201$~y. The extend of the x and y axes is $4.5\times 10^{17}$~cm,
$i=25^\circ$. Bottom right: [OIII]$\lambda5007$ long slit spectrum for
model~E taken after $t=1201$~y of evolution. The position axis extends
over $4.5\times 10^{17}$~cm, the velocity axis over 300~km~s$^{-1}$,
$i=25^\circ$. The diagram has been smoothed with a gaussian of
5~km~s$^{-1}$ FWHM.

\bye
\end